\documentstyle[12pt]{article}
\begin{document}
\title{ Vacuum polarization screening corrections to
the ground state energy of two-electron ions}
\author{A.N. Artemyev, V.M. Shabaev, and V.A. Yerokhin}

\date{}
\maketitle
\begin{abstract}
Vacuum polarization screening corrections to the ground state energy of
two-electron ions are calcualated in the range $Z=20-100$.
 The calculations are carried out for a finite nucleus charge distribution.
\end{abstract}
\section{Introduction}
It is well known that the dominant contribution to the vacuum polarization
correction in a strong Coulomb field arises from the Uehling
potential \cite{s1}. The remaining part of this correction is called by
the Wichman-Kroll contribution \cite{s2}.  Calculations of the
 Uehling contribution cause no problem and have been done
by many authors. Accurate calculations
of the Wichman-Kroll contribution were carried out in \cite{s3,s4,s5}.

Recent measurements of the Lamb shift in highly charged ions
\cite{s6,s7} have shown a necessity of calculation of  vacuum
polarization screening diagrams, i.e.  diagrams with one vacuum loop and
one electron-electron interaction (Fig.1).  In the present paper an
accurate calculation of these diagrams for the ground state of a
two-electron  ion is given.  Results of the calculations are compared with
a previous evaluation of this correction \cite{s8}.

Relativistic units ($\hbar =c=1$) are used in the paper.

\section{Basic formulas}

The vacuum polarization screening diagrams are
shown in Fig.1. The formal expressions for the energy level shift
due to these diagrams can easily be derived
using the two-time Green function method \cite{s9}.  The contribution of
the diagrams shown in Fig 1a is

\begin{eqnarray}
\Delta E_a=\sum\limits_P\left( -1\right)
^P\sum\limits_{\varepsilon _n\neq \varepsilon _a}\{\langle PaPb\mid I\left(
0\right) \mid nb\rangle \frac 1{\varepsilon _a-\varepsilon _n}\langle n\mid
U_{VP}^a\mid a\rangle + \nonumber \\
\langle PaPb\mid I\left( 0\right) \mid an\rangle \frac 1{\varepsilon
_a-\varepsilon _n}\langle n\mid U_{VP}^a\mid b\rangle + \nonumber \\
\langle Pa\mid U_{VP}^a\mid n\rangle \frac 1{\varepsilon _a-\varepsilon _n}
\langle nPb\mid I\left( 0\right) \mid ab\rangle + \nonumber \\
\langle Pb\mid U_{VP}^a\mid n\rangle \frac 1{\varepsilon _a-\varepsilon
_n}\langle Pan\mid I\left( 0\right) \mid ab\rangle \}.
\end{eqnarray}\\
Here $U_{VP}^a$ is the vacuum polarization potential:
\begin{eqnarray}
  U_{VP}^a({\bf x})&=&\frac{\alpha}{2\pi i}
\int d{\bf y}\frac{1}{|{\bf x}-{\bf y}|}
\int\limits_{-\infty }^\infty d\omega
Tr(G(\omega ,{\bf
y,y}))\,,\\
I(\omega,\mid {\bf x}-{\bf y\mid })&=&\alpha \frac{\alpha _{1\mu }\alpha
_2^\mu }{\mid {\bf x}-{\bf y\mid }}\exp (i\mid \omega \mid \mid {\bf
x}-{\bf y\mid )},
\end{eqnarray}
 $\alpha^{\mu}\equiv (1,\mbox{\boldmath $\alpha$})\,,\,\,$
$\mbox{\boldmath $\alpha$}$ are the Dirac matrices;
 $a$,  $b$ are the $1s$
states with a spin projection $m=\pm \frac {1}{2}$;
 $P$ is the permutation operator;
$G(\omega,{\bf x},{\bf y})=\sum\limits_n\frac{\psi ({\bf x})
\psi^{\dag}({\bf y})}{\omega -\varepsilon _{n}(1-i0)}$
is the Coulomb Green function.

The contribution of the
diagrams shown in Fig 1b is
\begin{eqnarray}
\Delta E_b=\sum\limits_P\left( -1\right) ^P\langle PaPb\mid U_{VP}^b
\mid ab\rangle,
\end{eqnarray}\\
where
\begin{eqnarray}
U_{VP}^b({\bf x,y})&=&\frac{\alpha^2}{2\pi i}\int\limits_{-\infty }^\infty
d\omega \int d {\bf z}_1\int d {\bf z}_2
\frac{\alpha_{1\mu}}{| {\bf x}-{\bf z}_1|}
\frac{\alpha_{2\nu}}{| {\bf y}-{\bf z}_2|}\nonumber \\ &&\times
Tr(\alpha^{\mu}G(\omega ,{\bf z}_1{\bf
,z}_2)\alpha^{\nu}G(\omega,{\bf z}_2{\bf ,z}_1))\,.
\end{eqnarray}
The contributions (1) and (4) are ultraviolet divergent. The most
simple way
to renormalize these contributions is to expand the vacuum loop in
powers of the external field ($\alpha Z$).  In this case the contribution
of the diagrams with an odd number of vertices in the vacuum loop (with
free-electron propagators) is equal to zero according to the Furry theorem.
In this expansion, only  the first nonzero term, called
as the Uehling term, is infinite. The charge
 renormalization  makes this
term finite and its calculation causes no problem.
The higher orders (in $\alpha Z$) terms  are
finite. However, regularization is still needed in the second
nonzero term due to the spurious gauge dependent piece of
the light-by-light scattering contribution.
As it was shown in \cite{s10,s11,s3}, in the calculation of
the vacuum polarization charge density, based on
the partial wave
 expansion of the electron Green function,
the spurious term does not contribute if
the sum over the angular momentum quantum number $\kappa$
 is restricted to a finite number of terms ($|\kappa|\leq K$).
 We found that this rule is also
correct in the case of the diagrams shown in Fig.1b (see Appendix
A).  Thus the Wichman-Kroll
contribution is calculated by summing up the partial differences between
the full contribution and the Uehling term.

\section{Calculation}

\subsection{Diagrams "a"}

Renormalized expression for the Uehling potential is well known
 \begin {eqnarray} U_{Uehl}^a(r)&=&-\alpha Z
\frac{2\alpha}{3\pi}\int\limits_0^\infty dr^{'}4\pi r'\rho
(r^{'})
\int\limits_1^\infty dt
(1 +\frac{1}{2t^2})
\frac{\sqrt{t^2-1}}{t^{2}}\nonumber \\
&&\times \frac{[\exp{(-2m|r-r'|t)}-\exp{(-2m(r+r')t)}]}
{4mrt} \,,
\end{eqnarray}
where $|e|Z\rho(r)$ is the density of the nucleus charge distribution
( $\int \rho(r) d{\bf r}=1$).
It is interesting to note that with a good precision
 ($\sim$ 0.3\% for $Z=80$) the formula (6) can be replaced
 by a simpler formula (see Appendix B):
\begin{eqnarray}
U_{Uehl}^a(r)\approx V(r)\frac{2\alpha }{3\pi
}\int\limits_1^\infty
dt ( 1+\frac
1{2t^2})\frac{\sqrt{t^2-1}}{t^{2}}
\exp (-2mrt)\,,
\end{eqnarray}
 where $V(r)$ is the potential of an extended nucleus.
The calculations were carried out
 for the Fermi model of the nuclear charge distribution
 using the exact formula (6).
To calculate the reduced Green
function which appears in the formula (1) the B-spline method
for the Dirac equation \cite{s12}  was used.
A change of the result, due to a one percent variation
of the root-mean-square charge radius of the nucleus,
was chosen as the uncertainty.

The calculation of the Wichman-Kroll contribution caused no
problem too.
Rotating
the contour of the $\omega$
integration in the complex $\omega$ plane along
the imaginary axis one get the following equation
\begin{eqnarray}
U_{WK}^a(x)&=&\frac{2\alpha }\pi \sum\limits_{\kappa =\pm 1 }^{\pm\infty}
\mid \kappa
\mid \int\limits_0^\infty d\omega \int\limits_0^\infty dy y^{2}
\int\limits_0^\infty dz z^{2}\frac {1}{max(x,y)}V(z)\nonumber \\
&&\times
\sum\limits_{i,k=1}^2
Re(F_{\kappa}^{ik}(i\omega,y,z)
(G_{\kappa}^{ik}(i\omega,y,z)- F_{\kappa}^{ik}(i\omega,y,z))),
\end{eqnarray}
where $G_{\kappa}^{ik}$
are the radial components of
the partial contributions to the bound-electron Green function
 and $F_{\kappa}^{ik}$
are the radial components of
the partial contributions to the free-electron Green function.
 The homogeneously charged
spherical shell model of the nucleus was used in the calculation
of $U_{WK}^{a}$.
In this case the Green function is expressed analytically
 in terms of the Whittaker,
 Bessel, and Hankel functions \cite{s10,s3}.
The reduced Green function which appears in (1) as well as
the $1s$ state wave function were calculated for the Fermi model.
 The  $\kappa$ series in (8) converges rapidly with
$\kappa$ increasing, so we needed $|\kappa| \leq 5$ to reach the relative
accuracy not worse than $10^{-4}$.

\subsection{Diagrams "b"}

The calculation of $\Delta E_b$ was carried out
 in the same way as  $\Delta E_a$.
The corresponding expression for the Uehling
 operator is also well known:
\begin{eqnarray}
U_{Uehl}^b({\bf x},{\bf y})=
\alpha \frac{\alpha _{1\mu}\alpha_2^\mu }{\mid
{\bf x}-{\bf y}\mid }
\frac{2\alpha}
{3\pi }\int\limits_1^\infty
dt ( 1+\frac
1{2t^2})\frac{\sqrt{t^2-1}}{t^{2}}
\exp (-2m\mid {\bf x}-{\bf
y}\mid t)
\,.
\end{eqnarray}
As before, the calculation of this term was made
 using the Fermi model of
the nucleus charge distribution. The uncertainty of the results was estimated
in the same way as for the Uehling part
 of the  diagrams "a".

The  calculation
of the Wichman-Kroll contribution of the diagram "b"
was done
 by summing  the partial
differences between the expression (5) and the corresponding
expression with the bound-electron Green functions  replaced by
the free-electron
Green functions. However, we encountered some obstacles trying to implement
this scheme directly. The origin of these obstacles is that,
 in the direct numerical calculation,
it is difficult  to reach a full cancelation of
some terms with large magnitude. To solve this problem we used
the Furry theorem.
Taking into account that according to this theorem
only terms containing even powers of $Z$ give
 nonzero contributions
to the diagram
shown in Fig 1b,
the Coulomb Green function was divided into two parts, each containing only
odd or even powers of the nucleus charge. After that the expression
$Tr(G_{k}G_{k})$ was replaced with
$Tr(G_{k}^{odd}G_{k}^{odd}+G_{k}^{even}G_{k}^{even})$.  Then the last
trace was calculated analytically (see Appendix C).  To control the accuracy
of this procedure  we calculated the
 correction due to interaction  with an  additional external field  $\Delta
V=-\frac \alpha r$ to the first order vacuum polarization contribution
for a point nucleus
(Fig.2).
In this case the sum of all the corrections must be equal to the
value $dE_{VP}/dZ$ where $E_{VP}$ is the first order vacuum polarization
contribution.
 The results of the
 test  are given in  the table 1.
The calculation of the Wichman-Kroll contribution of the diagram
shown in Fig.1b was carried out for a point nucleus.
However, because of the smallness of this contribution
($\sim 0.002$ eV for $Z$=90), the finite nuclear size correction
can be neglected.

The complete results of the calculation are presented in the table 2.
The values of the
root-mean-square charge radii were taken from \cite{s13}.
For comparison,  the results of a previous calculation
done in \cite{s8} are listed in the last column of the table.

\section*{Acknowledgements}

This work was supported in part by  Grant No. 95-02-05571a from
the Russian Foundation for Basic Research
and by personal (ANA) Grant No.  a314-f from International Soros Science
Educational Program.

\newpage
\appendix
\section{}
The spurious gauge dependent piece of the light-by-light scattering
contribution for the diagram shown in Fig.1b is given by
\begin{eqnarray}
\Delta E_{M}&=&e^{4}\sum_{P}(-1)^{P}\xi_{M}\,,\\
\xi_{M}&=&\frac{i}{2\pi}
\int_{C_F} d\omega
\int d{\bf x}_1 d{\bf x}_2 A_\mu^{(a)} ({\bf x}_1)
A_\nu^{(b)} ({\bf x}_2)
 \int d{\bf y}_1 d{\bf y}_2
\nonumber \\ &&\times
Tr[\alpha^\mu F_{M}(\omega,{\bf x}_1, {\bf y}_1)V({\bf y}_1)
F_{M}(\omega,{\bf y}_1, {\bf x}_2)\nonumber \\ &&\times
\alpha^\nu
F_{M}(\omega,{\bf x}_2, {\bf y}_2)V({\bf y}_2)F_{M}
(\omega,{\bf y}_2, {\bf x}_1)\nonumber \\ &&
+\alpha^\mu F_{M}(\omega,{\bf x}_1, {\bf x}_2)
\alpha^\nu
F_{M}(\omega,{\bf x}_2, {\bf y}_1)\nonumber \\ &&\times V({\bf y}_1)
F_{M}(\omega,{\bf y}_1, {\bf y}_2)V({\bf y}_2)F_{M}(\omega,{\bf y}_2, {\bf x}_1)
\nonumber \\ &&
+\alpha^\mu F_{M}(\omega,{\bf x}_1, {\bf y}_1)V({\bf y}_1)
F_{M}(\omega,{\bf y}_1, {\bf y}_2)\nonumber \\ &&\times
V({\bf y}_2)
F_{M}(\omega,{\bf y}_2, {\bf x}_2)\alpha^\nu F_{M}(\omega,{\bf x}_2, {\bf
x}_1)].
\end{eqnarray}
where $A_\mu^{(a)} ({\bf x})=\int d{\bf y} \psi_{Pa}^{\dag}({\bf y})
\alpha^\nu
D_{\nu\mu}(0,{\bf x}-{\bf y}) \psi_{a}({\bf y})$,
$D_{\nu\mu}(\omega,{\bf x}-{\bf y})$ is the photon propagator,
$F_{M}(\omega,{\bf x}, {\bf y})$ is the free electron propagator
with the electron mass replaced by a hypothetical heavy mass $M$.
Calculating a limit as $M$ becomes infinite yields
\begin{eqnarray}
\lim_{M\rightarrow\infty}
\xi_M&=&\xi^{C}+\xi^{B}\,,\nonumber \\
\xi^{C}&=&\frac{1}{4\pi i}\int_{C_F} d\omega \int d{\bf x}_{1} A_0^{(a)}
({\bf x}_{1})A_0^{(b)}({\bf x}_{1})V^2({\bf x}_{1})
\lim_{{\bf x}_{2}
\rightarrow
{\bf x}_{1}}\nonumber \\ &&\times
Tr\Bigl[\frac{d^3}{d\omega^3}F(\omega, {\bf x}_{1},{\bf x}_{2})\Bigr]\,,\\
\xi^{B}&=&\frac{i}{4\pi}\int_{C_F} d\omega \int d{\bf x}_{1} A_i^{(a)}
({\bf x}_{1})A_k^{(b)}({\bf x}_{1})V^2({\bf x}_{1})
\lim_{{\bf x}_{2}
\rightarrow
{\bf x}_{1}}\int d{\bf y}\nonumber \\ &&\times
Tr\Bigl[\alpha^{i}
\frac{d^2}{d\omega^2}
F(\omega, {\bf x}_{1},{\bf y})
\alpha^{k}
F(\omega, {\bf y},{\bf x}_{2})\Bigr]\,,
\end{eqnarray}
where $F(\omega, {\bf x}, {\bf y})$ is the free-electron propagator;
$i,k=1,2,3$.
The term $\xi^{C}$ can be treated in the same way as the
light-by-light scattering graph in the third-order vacuum
polarization \cite{s10,s3}. So, we restrict our consideration
to the term $\xi^{B}$. Using an identity
$\alpha^{k}=i[H_{F},x^{k}]$, where $H_{F}$ is the free-electron
Hamiltonian, we find
\begin{eqnarray}
\xi^{B}&=&\frac{1}{4\pi}\int_{C_F} d\omega \int d{\bf x}_{1}
 A_i^{(a)}({\bf
x}_{1})A_k^{(b)}({\bf x}_{1})V^2({\bf x}_{1})
\lim_{{\bf x}_{2}
 \rightarrow
{\bf x}_{1}}(x_{1}^k-x_{2}^k)\nonumber \\ &&\times
Tr(\alpha^i\frac{d^2}{d\omega^2}F(\omega, {\bf x}_{1},{\bf x}_{2}))
\nonumber \\
&=&\frac{1}{4\pi}\int d{\bf x}_{1} A_i^{(a)}({\bf x}_{1})A_k^{(b)}
({\bf x}_{1})V^2({\bf x}_{1})
\lim_{\Omega\rightarrow\infty}
\lim_{{\bf x}_{2} \rightarrow
{\bf x}_{1}}
(x_{1}^k-x_{2}^k)\nonumber \\ &&\times \left.Tr(\alpha^i\frac{d}{d\omega}
F(\omega, {\bf x}_{1},{\bf x}_{2}))\right|^{\omega=i\Omega}_{\omega=-i\Omega}.
\end{eqnarray}
The first derivative of $F$ with respect to $\omega$ is
\begin{eqnarray}
\frac{dF}{d\omega}=-\Bigl[\frac{d}{x}i\mbox{\boldmath$\alpha$}
\cdot{\bf x} +\beta m+\omega+\frac{d}{\omega x} \Bigr]
\frac{\exp{(-dx)}}{4\pi}
\frac{\omega}{d}\,,
\end{eqnarray}
where ${\bf x}={\bf x}_{1}-{\bf x}_{2}$, $x=|{\bf x}|$,
$d=\sqrt{m^2-\omega^2}$.
We can see that $\xi^{B}$ is equal to zero independently
of the order of taking limits in (A5). In the partial
expansion the terms $F_{\kappa}^{ik}(\omega, {\bf x}, {\bf y})$
are finite together with their first derivatives with respect
to $\omega$. It follows that the sum of a finite number of terms in
the $\kappa$ series is equal zero.

\section{}

Let us represent the Uehling
potential in the form:
\begin{eqnarray} U_{Uehl}^a(r)&=&-\alpha Z
\frac{2\alpha}{3\pi}\int\limits_1^\infty dt \frac{\pi}{mrt}
(1 +\frac{1}{2t^2})
\frac{\sqrt{t^2-1}}{t^{2}}\exp{(-2mrt)}A(r,t)\,,
\end{eqnarray}
where
\begin{eqnarray}
A(r,t)=\int_0^\infty dr'r' \rho(r') {[\exp{(-2m(|r-r'|-r)t)}-\exp{(-2mr't)}]}.
\end{eqnarray}
Expansion of the exponents in the Taylor series yields:
\begin{eqnarray}
{[\exp{(-2m(|r-r'|-r)t)}-\exp{(-2mr't)}]}= 4tmr_<+{\cal O}
(m^2r'r_<t^2).
\end{eqnarray}
Taking into account that
\begin{eqnarray}
\int_0^\infty dr'\rho(r')r'r_<=-\frac{r}{4\pi\alpha Z}V(r)
\end{eqnarray}
one can get
\begin{eqnarray}
U_{Uehl}^a(r)\approx V(r)\frac{2\alpha }{3\pi }\int\limits_1^\infty
dt (1+\frac 1{2t^2})\frac{\sqrt{t^2-1}}{t^{2}}
 \exp (-2mrt)\,.
\end{eqnarray}

\section{}
The analytical expressions for the radial components of the Coulomb Green
function are well known (see ,e.g.,[14]). In the units $\hbar=c=m=1$
for $x_{1} < x_{2}$
we have ($G=(\omega-H)^{-1}$)
\begin{eqnarray}
G^{11}_\kappa(\omega,x_1,x_2)&=&-(1+\omega)Q\left[(\lambda -
\nu)M_{\nu-\frac12,\lambda}(2dx_1) - (\kappa-\frac{\alpha Z}{d})
M_{\nu+\frac12,\lambda}(2dx_1)\right] \nonumber \\
&&\times \left[(\kappa+\frac{\alpha
 Z}{d})W_{\nu-\frac12,\lambda}(2dx_2)+W_{\nu+\frac12,\lambda}(2dx_2)\right]\,,
 \nonumber \\
 G^{12}_\kappa(\omega,x_1,x_2)&=&-dQ\left[(\lambda -
 \nu)M_{\nu-\frac12,\lambda}(2dx_1) - (\kappa-\frac{\alpha Z}{d})
 M_{\nu+\frac12,\lambda}(2dx_1)\right] \nonumber \\
&&\times  \left[(\kappa+\frac{\alpha
 Z}{d})W_{\nu-\frac12,\lambda}(2dx_2)-W_{\nu+\frac12,\lambda}(2dx_2)\right]\,,
 \nonumber \\
 G^{21}_\kappa(\omega,x_1,x_2)&=&-dQ\left[(\lambda -
 \nu)M_{\nu-\frac12,\lambda}(2dx_1) + (\kappa-\frac{\alpha Z}{d})
 M_{\nu+\frac12,\lambda}(2dx_1)\right]\,,
 \nonumber \\
&&\times  \left[(\kappa+\frac{\alpha
 Z}{d})W_{\nu-\frac12,\lambda}(2dx_2)+W_{\nu+\frac12,\lambda}(2dx_2)\right]\,,
 \nonumber \\
 G^{22}_\kappa(\omega,x_1,x_2)&=&-(1-\omega)Q\left[(\lambda -
 \nu)M_{\nu-\frac12,\lambda}(2dx_1) + (\kappa-\frac{\alpha Z}{d})
 M_{\nu+\frac12,\lambda}(2dx_1)\right] \nonumber \\
&&\times  \left[(\kappa+\frac{\alpha
 Z}{d})W_{\nu-\frac12,\lambda}(2dx_2)-W_{\nu+\frac12,\lambda}(2dx_2)\right]\,,
 \end{eqnarray}
 where $d=\sqrt{1-\omega^2}$, $\lambda=(\kappa^2-(\alpha
 Z)^2)$, $\nu=\frac{\alpha Z \omega}{d}$, $Q=\frac{1}{4d^2(x_1x_2)^\frac32}
\frac{\Gamma(\lambda-\nu)}{\Gamma(1+2\lambda)}$, $M_{\alpha,\beta}$, and
$W_{\alpha,\beta}$ are the Whittaker functions.
For $x_{1}>x_{2}$ the radial Green functions can be obtained from
the symmetry condition
$$
G_{\kappa}^{ik}(\omega,x_{1},x_{2})
=G_{\kappa}^{ki}(\omega,x_{2},x_{1})\,.
$$
Defining
\begin{eqnarray}
A&=&-Q(\lambda-\nu)M_{\nu-\frac12,\lambda}(2dx_1)W_{\nu-\frac12,\lambda}(2dx
_2)\,,
\nonumber \\
B&=&-Q(\lambda-\nu)M_{\nu-\frac12,\lambda}(2dx_1)W_{\nu+\frac12,\lambda}(2dx
_1)\,,
\nonumber \\
C&=&-QM_{\nu+\frac12,\lambda}(2dx_1)W_{\nu-\frac12,\lambda}(2dx_2)\,,
\nonumber \\
D&=&-QM_{\nu+\frac12,\lambda}(2dx_1)W_{\nu+\frac12,\lambda}(2dx_2)\,,
\end{eqnarray}
and taking into account that for $\omega=i\varepsilon$, where $\varepsilon$ is
real,  changing the sign of $Z$ is equal to
replacing $A$, $B$, $C$, and $D$ with their complex conjugated values,
we find for the radial parts of the Coulomb Green function
containing only odd or even powers of $Z$ the following expressions
\begin{eqnarray}
G^{11,odd}_\kappa&=&(1+\omega) \bigl \{ i(\kappa Im(A-D)+Im(B)-\gamma
Im(C))\nonumber \\ &&+ \frac{\alpha Z}{d}Re(A+D)\bigr \}\,, \nonumber\\
G^{12,odd}_\kappa&=&d
\bigl \{ i(\kappa Im(A+D)-Im(B)-\gamma Im(C))+ \frac{\alpha
Z}{d}Re(A-D)\bigr \}\,, \nonumber\\
G^{21,odd}_\kappa&=&d \bigl \{ i(\kappa
Im(A+D)+Im(B)+\gamma Im(C))+ \frac{\alpha Z}{d}Re(A-D)\bigr \}\,,\nonumber\\
G^{22,odd}_\kappa&=&(1-\omega) \bigl \{ i(\kappa
Im(A-D)-Im(B)+\gamma Im(C)) \nonumber \\ &&
+ \frac{\alpha Z}{d}Re(A+D)\bigr \}\,,\\
G^{11,even}_\kappa&=&(1+\omega) \bigl \{ \kappa Re(A-D)+Re(B)-\gamma
Re(C)\nonumber \\ &&+ i\frac{\alpha Z}{d}Im(A+D)\bigr \}\,, \nonumber \\
G^{12,even}_\kappa&=&d \bigl \{ \kappa Re(A+D)-Re(B)-\gamma
Re(C)+ i\frac{\alpha Z}{d}Im(A-D)\bigr \}\,, \nonumber \\
G^{21,even}_\kappa&=&d \bigl \{ \kappa Re(A+D)+Re(B)+\gamma
Re(C)+ i\frac{\alpha Z}{d}Im(A-D)\bigr \}\,, \nonumber\\
G^{22,even}_\kappa&=&(1-\omega) \bigl \{ \kappa Re(A-D)-Re(B)+\gamma
Re(C)\nonumber \\ &&+ i\frac{\alpha Z}{d}Im(A+D)\bigr \}\, ,
\end{eqnarray}
where $\gamma=\kappa^2-\frac{(\alpha Z)^2}{d^2}$.

In the calculation of the Wichman-Kroll part of the diagram shown
in Fig.1b the
expression $Re(Tr(\alpha^\nu G(\omega,x,y) \alpha^\mu G(\omega,y,x)))$
 appears.
After having integrated over the
angles we have to evaluate the following expressions
\begin{eqnarray}
S_0&=&Re\bigl(\sum_{sign(\kappa)=-1}^1 \sum_{i,k=1}^2
(G^{ik}_\kappa(\omega,x,y))^2\bigr)\,,\\
S_1&=&Re\bigl(\sum_{sign(\kappa)=-1}^1(G^{11}_{\kappa}
G^{22}_{\kappa}+G^{22}_{\kappa}G^{11}_{\kappa}+
G^{12}_{\kappa}G^{21}_{\kappa}+G^{21}_{\kappa}G^{12}_{\kappa})\bigr)\,\\
S_2&=&2Re\bigl(\sum_{sign(\kappa)=-1}^1
(G^{11}_{\kappa}
G^{22}_{\kappa'}+G^{22}_{\kappa}G^{11}_{\kappa'}+
G^{12}_{\kappa}G^{21}_{\kappa'}+G^{21}_{\kappa}G^{12}_{\kappa'})\bigr)\,,
\end{eqnarray}
where $\kappa'=-sign(\kappa)(|\kappa|+1)$.
Using (C3)-(C4) and the Furry theorem
 we transformed these expressions to the following ones
\begin{eqnarray}
S_0&=& 8\bigl \{ \bigl(\frac{(\alpha
Z)^2}{d^2}+\kappa^2\bigr)\bigl(Re(A)^2+Re(D)^2-Im(A)^2-Im(D)^2\bigr)
\nonumber \\ &&+Re(B)^2 - Im(B)^2 + \gamma \bigl(Re(C)^2-Im(C)^2\bigr)
\bigr \} \nonumber\\ &&-16\varepsilon \frac{(\alpha Z)}{d}\bigl \{ Re(A+D)
\bigl(Im(B)-\gamma Im(C)\bigr)\nonumber \\ &&
+Im(A+D)\bigl(Re(B)-\gamma Re(C)\bigr)\bigr \} \nonumber \\
&&-16\varepsilon^2 \bigl \{ (\frac{(\alpha Z)^2}{d^2}+\kappa^2)\bigl(
Re(A)Re(D)+Im(A)Im(D)\bigr)
\nonumber \\ && +\gamma \bigl(Re(B)Re(C)+Im(B)Im(C)\bigr) \bigr
\}\\
S_{1}&=&8d^2\bigl \{ \bigl(\kappa^2+\frac{(\alpha
Z)^2}{d^2}\bigr)\bigl(Re(A)^2+Re(D)^2- Im(A)^2-Im(D)^2\bigr)
\nonumber \\ &&+\gamma
\bigl(Im(C)^2-Re(C)^2\bigr)+Im(B)^2-Re(B)^2 \bigr \}\\
S_{2}&=&16d^2\bigl
\{ \bigl(|\kappa|(|\kappa|+1)-\frac{(\alpha Z)^2}{d^2}\bigr)
\nonumber \\ &&\times
\bigl(Im(A)Im(A')+Im(D)Im(D')- Re(A)Re(A')-Re(D)Re(D')\bigr)\nonumber \\ &&
+Im(B)Im(B')-Re(B)Re(B')+
\gamma \gamma'\bigl(Im(C)Im(C')\nonumber \\ &&-
Re(C)Re(C')\bigr)\bigr \}
\end{eqnarray}
The equations (28)-(30) were used in the numerical calculation.

\newpage
\begin{table}
\caption{Control calculation of the correction due to interaction
with an additional external field $\Delta V(r)=-\alpha/r$ to
the first order vacuum polarization contribution for the $1s$ state
 (see Fig.2) (in eV). The label "a" corresponds to the sum
of the first two diagrams in Fig 2. The label "b" corresponds
to the third diagram in Fig 2.
 The calculation is carried out for a point
nucleus.}

\begin{tabular}{|c|l|l|l|l|l|l|}\hline
$Z$ & $\Delta E_{Uehl}^a$ &$\Delta E_{Uehl}^b$ & $\Delta E_{WK}^a$
& $\Delta E_{WK}^b$ &$\Delta E_{VP}$
& $ \frac{d E_{VP}}{dZ}$ \\ \hline

20&$-0.019183$&$-0.006487$&$0.000072$&$0.000076$&$0.025522$&$0.025521$
\\ \hline
40&$-0.16181$&$-0.05215$&$0.00206$&$0.00221$&$0.20970$&$0.20969$\\ \hline
60&$-0.67136$&$-0.19446$&$0.01650$&$0.01702$&$0.83231$&$0.83234$\\ \hline
90&$-4.5280$&$-0.9586$&$0.1970$&$0.1711$&$5.1185$&$5.1186$\\ \hline
100&$-9.2613$&$-1.6460$&$0.4532$&$0.3543$&$10.1000$&$10.0999$\\ \hline
\end{tabular}
\end{table}
\newpage
 \begin{table}
 \caption{Contribution of  the vacuum polarization screening diagrams
 to the ground state
energy of two-electron ions (in eV).}
 \begin{tabular}{|c|l|l|l|l|l|l|c|}\hline

$Z$&$<r^2>^{1/2}$ (fm)&$\Delta E_{Uehl}^a$&$\Delta E_{Uehl}^b$&
$\Delta E_{WK}^a$&$\Delta E_{WK}^b$&$\Delta E_{VP}$ & Ref. [8] \\ \hline

20&$3.478$&$0.0090$&$0.0010$&$-0.0000$&$-0.0000$&$0.0100$& \\ \hline

30&$3.928$&$0.0316$&$0.0035$&$-0.0003$&$-0.0000$&$0.0348$& \\ \hline

32&$4.072$&$0.0388$&$0.0043$&$-0.0004$&$-0.0000$&$0.0427$&0.0 \\ \hline

40&$4.270$&$0.0807$&$0.0091$&$-0.0011$&$-0.0000$&$0.0887$& \\ \hline

50&$4.655$&$0.176$&$0.020$&$-0.003$&$-0.000$&$0.192$& \\ \hline

54&$4.787$&$0.234$&$0.027$&$-0.005$&$-0.000$&$0.255$&0.2 \\ \hline

60&$4.914$&$0.350$&$0.040$&$-0.009$&$-0.000$&$0.380$& \\ \hline

66&$5.224$&$0.515$&$0.058$&$-0.016$&$-0.000$&$0.557$&0.6 \\ \hline

70&$5.317$&$0.661$&$0.074$&$-0.022$&$-0.000$&$0.713$& \\ \hline

74&$5.373$&$0.845$&$0.093$&$-0.030$&$-0.000$&$0.908$&0.9 \\ \hline

80&$5.467$&$1.215$&$0.132$&$-0.049$&$\;\;\;0.000$&$1.298$& \\ \hline

83&$5.533$&$1.455$&$0.156$&$-0.062$&$\;\;\;0.001$&$1.550$&1.6 \\ \hline

90&$5.645$&$2.214(1)$&$0.230$&$-0.105$&$\;\;\;0.002$&$2.341(1)$& \\ \hline

92&$5.860$&$2.493(2)$&$0.256$&$-0.122$&$\;\;\;0.003$&$2.630(2)$&2.6 \\ \hline

100&$5.886$&$4.063(4)$&$0.398$&$-0.221$&$\;\;\;0.008$&$4.248(4)$& \\ \hline
\end {tabular}
\end {table}
\newpage
\newcommand{\electronline}{\line(0,1){100}}
\newcommand{\photonlineleft}{
 \multiput(0,0)(-8,0){5}{\oval(4,4)[b]}
 \multiput(-4,0)(-8,0){4}{\oval(4,4)[t]}}
\newcommand{\photonlineright}{
\multiput(0,0)(8,0){5}{\oval(4,4)[b]}
\multiput(4,0)(8,0){4}{\oval(4,4)[t]}}

\newcommand{\photonliner}{
\multiput(0,0)(8,0){2}{\oval(4,4)[b]}
\multiput(4,0)(8,0){1}{\oval(4,4)[t]}}

\newcommand{\photonl}{
\multiput(0,0)(8,0){2}{\oval(4,4)[b]}
\multiput(4,0)(8,0){2}{\oval(4,4)[t]}}

\newcommand{\photonlinel}{
 \multiput(0,0)(-8,0){2}{\oval(4,4)[b]}
 \multiput(-4,0)(-8,0){1}{\oval(4,4)[t]}}

\newpage

\pagestyle{empty}
\begin{figure}
\caption{Vacuum polarization screening diagrams}
   \begin{picture}(480,400)
    \put(135,65){
      \put(63,0){\electronline}
      \put(61,50){\photonlinel}
	  \put(45,50){\circle {12}}
      \put(29,50){\photonliner}
     \put(27,0){\electronline}
 }
\put(180,45){b}

   \put(75,210){
      \put(63,0){\electronline}
      \put(61,75){\photonlineleft}
      \put(61,25){\photonlinel}
	  \put(45,25){\circle {12}}
     \put(27,0){\electronline}
 }\put(180,190){a}

    \put(0,210){
      \put(42,0){\electronline}
      \put(44,75){\photonlineright}
      \put(44,25){\photonliner}
	  \put(60,25){\circle {12}}
     \put(78,0){\electronline}
          }
   \put(260,210){
      \put(63,0){\electronline}
      \put(61,25){\photonlineleft}
      \put(61,75){\photonlinel}
	  \put(45,75){\circle {12}}
     \put(27,0){\electronline}
    }
    \put(185,210){
      \put(42,0){\electronline}
      \put(44,25){\photonlineright}
      \put(44,75){\photonliner}
	  \put(60,75){\circle {12}}
\put(78,0){\electronline}

    }
  \end{picture}

\end{figure}
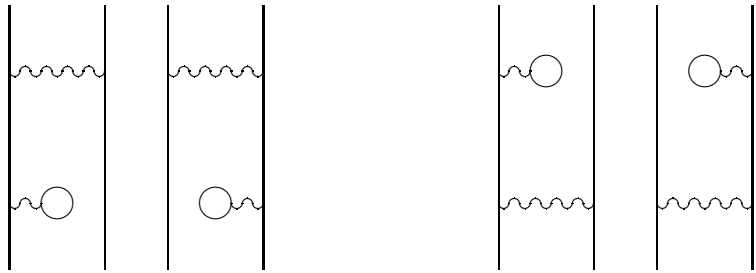
\newpage
\begin {figure}
\caption{Diagram equation for control calculation of the correction
due to interaction with an
external field $\Delta V=-\alpha/r$ to the first order vacuum polarization
contribution.}

   \begin{picture}(480,400)

    \put(0,210){
      \put(42,0){\electronline}
      \put(44,50){\photonl}
	  \put(44,25) {\line(1,0){10}}
	  \put(64,25) {\line(1,0){10}}
	  \put(69,30) {\line(1,-1){10}}
	  \put(69,20) {\line(1,1){10}}
	  \put(68,50){\circle {19}}
          }
    \put(90,210){
      \put(42,0){\electronline}
      \put(44,50){\photonl}
	  \put(44,75) {\line(1,0){10}}
	  \put(64,75) {\line(1,0){10}}
	  \put(69,80) {\line(1,-1){10}}
	  \put(69,70) {\line(1,1){10}}
	  \put(68,50){\circle {19}}
          }
    \put(180,210){
      \put(42,0){\electronline}
      \put(44,50){\photonl}
	  \put(80,50) {\line(1,0){10}}
	  \put(85,55) {\line(1,-1){10}}
	  \put(85,45) {\line(1,1){10}}
	  \put(68,50){\circle {19}}
          }
    \put(310,210){
      \put(42,0){\electronline}
      \put(44,50){\photonl}
	  \put(68,50){\circle {19}}
          }
	\put(100,260){+}
	\put(190,260){+}
	\put(300,260){= {\huge $\frac {d}{dZ}$}}
  \end{picture}
\end {figure}
\end{document}